\documentstyle[11pt,aaspp]{article}
\input epsf
\def\l*{$L_*$\/}
\def\etal{{\it et al. }}

\def\kms{\rm ~km~s^{-1}}

\def\kmsmpc{km s$^{-1}$ Mpc$^{-1}$\ }

\def\l*{$L_{*}$}

\def\gsim{ \lower .75ex \hbox{$\sim$} \llap{\raise .27ex \hbox{$>$}} }
\def\lsim{ \lower .75ex \hbox{$\sim$} \llap{\raise .27ex \hbox{$<$}} }
\def\pp{\noindent\parshape 2 0truecm 16.0truecm 2.0truecm 15truecm}

\def\spose#1{\hbox to 0pt{#1\hss}}
\def\simlt{\mathrel{\spose{\lower 3pt\hbox{$\mathchar"218$}}
     \raise 2.0pt\hbox{$\mathchar"13C$}}}
\def\simgt{\mathrel{\spose{\lower 3pt\hbox{$\mathchar"218$}}
'     \raise 2.0pt\hbox{$\mathchar"13E$}}}

\input{epsf}

\begin{document}

\title{Resolving the Structure of Cold Dark Matter Halos}

\vskip 0.5truecm

\centerline{\bf Moore B.$^{\bf 1}$,
Governato, F.$^{\bf 1}$,
Quinn T.$^{\bf 2}$,
Stadel J.$^{\bf 2}$
\& Lake G.$^{\bf 2}$}

$^{\bf 1}$ {Department of Physics, University of Durham, 
Durham City, DH1 3LE, UK}

$^{\bf 2}$ {Department of Astronomy, University of Washington, 
Seattle, WA 98195, USA}

\begin{abstract}

We examine the effects of mass resolution and force softening on the
density profiles of cold dark matter halos that form within
cosmological N-body simulations.  As we increase the mass and force
resolution, we resolve progenitor halos that collapse at higher
redshifts and have very high densities.  At our highest resolution we
have nearly 3 million particles within the virial radius, several
orders of magnitude more than previously used and we can resolve more
than one thousand surviving dark matter halos within this
single virialised system.  
The halo profiles become steeper in the central regions and we may not
have achieved convergence to a unique slope within the inner 10\% of
the virialised region. Results from two very high resolution halo
simulations yield steep inner density profiles, $\rho(r)\sim
r^{-1.4}$.  The abundance and properties of arcs formed within this
potential will be different from calculations based on lower
resolution simulations. The kinematics of disks within such a steep
potential may prove problematic for the CDM model when compared
with the observed properties of halos on galactic scales.

\end{abstract}

\vfil\eject

\section{Introduction}

The cold dark matter (CDM) model is a highly successful and well
motivated cosmological model.  The basic premises are an inflationary
universe dominated by a dark matter particle, such as the axion, that
leads to ``bottom up'' hierarchical structure formation. Small dense
halos collapse at high redshifts and merge successively into the large
virialised systems that contain the galaxies that we observe today
({\it e.g.} Davis \etal 1985 and references therein).  On scales
$\gsim $ several Mpc, variants from the standard 
model can successfully reproduce the observed
clustering pattern of galaxies together with the abundances of
clusters \--- two fundamental statistics that are sensitive to both
the shape and amplitude of the power spectrum (Eke \etal
1996). Comparisons between the model and data on smaller scales test a
different aspect of the CDM paradigm. For example, the internal
structure of halos can be compared with the dynamics of dwarf
galaxies. This test shows that the dark matter is not as concentrated
as numerical simulations of CDM indicate, perhaps indicating that the
dark matter is not as ``cold'' as CDM (Moore 1996, Flores \etal 1996).

The properties of dark matter halos in the CDM model have been
extensively investigated by numerous researchers since the early 80's,
however, only in the past decade have computational facilities and
software improved to the point such that the central properties of
dark halos can be compared directly with the observations. 
The first results of Zurek \etal (1986) and Frenk \etal (1988) did not
have the resolution necessary to probe the ``observed'' region of dark
halos, but did show isothermal structures in the outer regions.
Higher resolution simulations by Dubinski \& Carlberg (1991), Warren
\etal (1992), Carlberg (1994), Crone \etal (1994) showed evidence
for density profiles with slopes flatter than isothermal in the
central regions, with densities varying as $r^{-1}$ in the inner 10\%
of the halos. Navarro, Frenk \& White (1996, hereafter NFW) made a
systematic study of CDM halo structure over a range of mass
scales. They found that the density profiles of halos follow a
universal form, uniquely determined by their mass and virial radius;
varying from $r^{-1}$ in the central regions, smoothly rolling over to
$r^{-3}$ at the virial radii. (The virial radius, $r_{vir}$, is
defined as the radius of a sphere containing a mean mass
over-density of 200 with respect to the global value). The NFW halos
typically contained 5,000 \--- 10,000 particles, a number that was
claimed to be sufficient to resolve the density profile of halos 
beyond a distance $\sim 1\%$ of the virial radius. 

In this paper we investigate the inner structure of CDM halos using
N-body simulations with much higher mass and force resolution than
used previously.  We shall test separately the effects of softening
and particle number on the final density profiles and the issue of
convergence to a unique profile. Similar tests were performed by
Tormen \etal (1996) and Craig (1997). Although the scale of their
tests were smaller than carried out here, their results suggested a
similar dependence on numerical resolution.  We begin with a brief
discussion of the numerical techniques and simulation parameters.  In
Section 3 we show visually and graphically how the structure and
substructure of dark halos changes with numerical resolution and we
attempt to separate the affects of force softening and mass resolution
on the density profiles.  Finally, we discuss the role of resolution
in determining halo density profiles and the possible origin of the
steep inner density cusp that we find.

\section{Techniques and initial conditions}
 
Simulating the formation of a dark matter halo requires an accurate
treatment of the cosmological tidal field from a large volume of the
universe, whilst obtaining high resolution throughout the infall and
turnaround regions that will constitute the final halo.  A candidate
halo is initially identified from a large cosmological volume that has
been simulated at lower resolution. The particles within the selected
halo are traced back to the initial conditions to identify the region
that will be re-simulated at higher resolution. The power spectrum is
extrapolated down to smaller scales, matched at the boundaries such
that both the power and waves of the new density field are identical
in the region of overlap, then this region is populated with a new
subset of less massive particles. Beyond the high resolution region
the mass resolution is decreased in a series of shells such that the
external tidal field is modelled correctly in a cosmological
context. The starting redshift is increased such that the initial
fluctuations are less than one percent of the mean density
and we then re-run the simulation to the present epoch.

The particle distribution is evolved using a new high performance
parallel treecode, ``PKDGRAV'', that has accurate periodic boundaries
(previous studies using treecodes had vacuum boundary conditions), and
a variable timestep criteria based upon the local acceleration.  The
code uses a co-moving spline softening length such that the force is
completely Newtonian at twice our quoted softening lengths. In terms
of where the force is 50\% of the Newtonian force, the equivalent
Plummer softening length would be 0.67 times the spline softening
length.  We resimulate two
halos, both extracted from standard CDM simulations normalised such
that $\sigma_8=0.7$ and the shape parameter $\Gamma=0.5$ (H=50\kmsmpc\
is adopted throughout).  The ``Virgo'' halo has a virial radius of 2
Mpc and was extracted from an initial simulation with a box length of
100 Mpc. The ``Coma'' halo has a virial radius of 3.4 Mpc and was
extracted from a larger initial simulation that had a volume of one
Gpc$^3$. We performed several simulations of each halo; at our highest
resolution the particle mass in the highest resolution regions was
$8\times10^8M_\odot$ and we used a 5 kpc softening length.

PKDGRAV has been extensively tested using a variety of methods, for
example, comparing forces with those calculated using a direct $N^2$
code on a clustered particle distribution or comparing the growth of
waves against linear predictions. We have made one additional test by
simulating a halo with PKDGRAV and a P$^3$M code (Couchman 1991).  We
used the ``cluster comparison'' data generated by (Frenk \etal 1998).
This is constrained realization of a standard $\Omega=1$ cold dark
matter universe such that a rich cluster forms at the center of a 64
Mpc box.
The cluster contains over $10^5$ particles within $r_{vir}$ at the
final time and both simulations use a force softening with a
resolution of $0.5\%r_{vir}$.  The global evolution was virtually
identical in each simulation; the final virial radii are within 0.3\%
of each other and the density profiles agreed extremely well on all
scales.

\section{Results}

Colour Plate 1 shows the final density distribution of our highest
resolution simulation of the ``Coma'' cluster.  The colour scale in
both panels represents the local over-density at the position of each
particle and reflects a variation from $10^1$ to $10^6$ times the mean
density. The upper panel shows the entire 1000 Mpc box and illustrates
the nested refinement zones, although plotted on this scale the
central cluster is not resolved.  In the lower panel we have extracted
a sphere of radius 3.4 Mpc = $r_{vir}$ around the cluster.  A wealth
of substructure and ``halos within halos'' can be seen within this
cluster.



The left and right hand panel of Colour Plate 2 shows the physical
densities and the phase space densities of the Coma halo simulated at
three different resolutions.  These quantities are plotted at the
position of each particle using a colour scale spanning 5 orders of
magnitude and they are measured by smoothing over the nearest 64
particles.  The upper plot shows the cluster simulated at a resolution
similar to that used by NFW.  There are about 13,000 particles within
the virial region. The central and lower plots show the same cluster
simulated at higher resolution such that there are $1\times 10^5$ and
$2.7\times 10^6$ particles respectively within the virial radius.  The
visual difference is remarkable.  As we increase the mass resolution
we begin to resolve more and more substructure; at the highest resolution
there are $\sim 1500$
halos within $r_{vir}$.  A detailed analysis of the properties and
dynamics of the dark matter substructure within the ``Virgo'' cluster
simulation can be found in Ghigna \etal (1998). {\footnote{We
note that the inner 10\% of even the highest resolution cluster is
entirely smooth. The overmerging problem is still inherent in
dark matter simulations as a result of halo\--halo collisions and
the global tidal field destroying halos that pass close to 
the center (Moore, Katz \& Lake 1996).}}

In Figure 1 we show the density profile (calculated in spherical
shells) of the ``Coma'' cluster simulated with different mass and
force resolutions.  The halo centers are identified using both the
center of mass of the inner halo and using the most bound particle and
both methods give very similar results. We find that at higher
resolutions, the central halo density profiles becomes steeper and as
we increase the mass resolution by a factor of 200, the density at 1\%
of the virial radius increases by 250\%.


We have performed 5 simulations of this cluster in which we
systematically increased the mass resolution such that we have a total of
92, 1450, 13000, 100000 and 2700000 particles within $r_{vir}$.
In each simulation we set the softening parameter to be 1/50'th of
the mean inter-particle separation in the highest resolution region;
in the highest resolution run, the softening was equal to 5 kpc.
For each simulation we determine the radius at which the 
density profile
falls by 20\% with respect to the highest resolution simulation and
define this to be a ``reliable'' radius beyond which we can trust the
results. We find that this radius
decreases systematically from 13\%, 4.5\%, 2.8\%, 1.9\% of $r_{vir}$
and the total number of particles that lie within these
radii are 9, 35, 160, 700 respectively.  We
conclude from this experiment that setting a minimum particle number
is an insufficient criterion for defining the radius outside of which
to ``believe'' the density profile. In fact, the reliable radius
defined in this way is very close to a scale equal to the 0.5 times 
the mean inter-particle separation of those particles within $r_{vir}$.


These convergence tests show that a factor of 8 in mass resolution
leads to a factor of 4 increase in the number of particles needed to
define the reliable radius. We therefore estimate that the highest
resolution run is correct to a radius that contains about 5600
particles which is 0.8\% of $r_{vir}$. At this point the slope of the
density profile has converged to a value of $r^{-1.4}$. It is possible
that further increases in resolution will trace out this asymptotic
gradient to higher densities but not to significantly steeper slopes.

\subsection{Force softening and particle number}

We are attempting to model a collisionless system by sampling phase 
space with a small number of particles. This sampling introduces numerical
noise that is ameliorated by softening the gravitational force on 
small scales that can effect the halo properties. The
number of particles defines the smallest halo that can collapse at
high redshifts and thus introduces a maximum phase space density that
can be resolved. The effect of softening is to introduce a ``soft''
core at the centers of all the halos approximately the 
size of the softening length. This causes halos to be disrupted more 
easily from tidal forces \--- the mechanism responsible for the 
over-merging problem (Moore, Katz \& Lake 1996).

At a fixed mass resolution, decreasing the softening will lead to
higher central densities, but there will be a point at which smaller
softening lengths will not change the inner profile. At this point the
density profile turns over to a constant value that is set by the
maximum density that can be resolved by the first collapsing halos
\--- unless artificial relaxation occurs.  Figure 2 shows the effects
of softening on the ``Virgo'' cluster profile. This halo has $\sim
20,000$ particles within the virial radius of 2 Mpc and we simulate it
using values of the softening length ranging from 1 kpc to 1
Mpc. The density profiles are plotted on scales smaller than the
adopted softening lengths to indicate the behaviour in the inner
regions, but results on scales less than the softening length
have no physical meaning.
Having said this, we note how well the density
profile agrees on all scales slightly larger than the softening
length. As the softening length is decreased we converge upon the same
profile found by NFW, with a slope tending towards $r^{-1}$ in the
inner regions.  Eventually we run out of particles to resolve the
inner regions and decreasing the softening from 20 kpc to 10 kpc has
very little effect on the halo profile. {\footnote{Although not
plotted here, values of the co-moving 
softening below 5 kpc lead to visible signs of relaxation in 
the density profile \-- the central region begins to expand. 
This evidence for collisionality occurs for values of the softening
smaller than 1/200'th of the mean inter-particle separation within 
the cosmological volume simulated and less than 1/20'th 
the mean inter-particle separation within the virial radius.
}}

The solid lines in Figure 2 show the same cluster simulated with 10
kpc softening, but with 20 times as many particles. This leads to a
significantly steeper density profile than the same halo simulated
with the same softening but using fewer particles. At a radius equal
to 1\% of the virial radius, just over two softening lengths, the
density increases by 70\%. Within this radius the low resolution run
contains over 50 particles.  The two solid lines in Figure 2 show the
density profiles of the same halo at redshifts z=0 and z=0.25, several
billion years before the final output.  This halo contains almost half
a million particles within the virial radius and over 90\% of the mass is
in place at a redshift z=0.25. The density profiles normalised to
$r_{vir}$ are almost identical at both epochs, demonstrating that
relaxation at late times or other numerical 
artifacts are not affecting our results
even in the central regions.

\subsection{Fundamental scatter in the density profiles}

Although a great deal of substructure is evident in these simulations,
the bulk of the mass distribution in these halos lies within a smooth
background of particles that were tidally stripped from the infalling
halos.  A single subclump with velocity dispersion $\sigma_{sub}$ will
introduce a fluctuation in the density profile
$\delta \rho/\rho \sim \delta M_{sub}/\delta M_{clus} \sim
(\sigma_{sub}/\sigma_{clus})^2.$  Therefore, even the largest clumps of
substructure with circular velocity $\sim 300 \kms$ will only produce
fluctuations of order 5\% in a rich, virialised, galaxy cluster.  The global
profiles may also vary from cluster to cluster, perhaps as a result of
different formation histories.



The residuals from the best fit NFW profiles to these halos show a
characteristic ``S'' shape with deviations of order 20\%.  Within 10\%
of the virial radius a power law of slope ($\sim r^{-1.4}$) is an
excellent fit. A simple modification to the NFW profile from
$\rho(R)\propto(R(1+R^2))^{-1}$ to
$\rho(R)\propto(R^{1.4}(1+R^{1.4}))^{-1}$ fits our high resolution
simulations very well.  (Here R is the radius from the cluster center
expressed in terms of the virial radius divided by a scale factor that
is equal to $r_{vir}\times0.18$ for the halos simulated here).  We
emphasize the need to perform many more halo simulations at this
resolution to study the scatter in the inner and outer slopes and the
concentration parameter.

\section{Achieving convergence in halo properties \--- what's going on?}

What are the possible numerical or physical reasons behind the shape
of CDM density profiles and the dependence upon resolution
that we have investigated here?

The simplest interpretation is that the gravitational softening is
affecting our results on scales less than about 4 or 5 softening
lengths; all of the halo density profiles we have calculated agree
very well beyond this scale.  Thus, at our highest resolution we have
converged upon a unique solution and we can resolve the density
profiles to $\sim 1\%$ of $R_{vir}$ with $10^6$ particles and softening 0.2\%
$R_{vir}$. However, we have demonstrated that changing the softening
whilst keeping the mass resolution fixed leads to agreement between
profiles on scales just larger than one softening length.

Splinter \etal (1997 and references within) 
emphasize that many statistics are not
accurately reproduced using N-body codes that are intrinsically
collisional. In comparison with PM codes, they argue that only
scales larger than the mean inter-particle separation should be 
considered. This would correspond to a scale about 10\% $r_{vir}$
within our highest resolution simulation \-- an order of magnitude
larger than our convergence tests suggest.

Evans \& Colett (1997) demonstrated that a density profile with
a slope of $-4/3$ is a stable solution to the Fokker-Planck
and the collisional Boltzmann equations. However, we note that the trend
in Figure 1 is the opposite as expected from collisional effects.
Furthermore, we do not observe a significant inwards 
energy transfer as expected from the Evans \& Colett model.

The inner density profiles may hold clues as to the initial power
spectrum.  For example, Subramanian \& Ostriker (1998) extended
Bertschinger's (1985) self-similar solutions to the spherical collapse
model to include angular-momentum. They find that the slope of the profiles
are directly related to the degree of translational motion and the
spectrum of density fluctuations.

Steeper profiles may arise from substructure halos
that are transferring material at high phase space density to the
center of the cluster (Syer \& White 1997). 
As we move to higher resolution we can resolve
structure collapsing at earlier redshifts. This material at the
highest physical density, invariably ends up at the center of the
cluster by $z=0$. Because it is denser it is more robust to tidal
disruption by the cluster, therefore has the potential of carrying
more mass to the central regions.

We demonstrate this latter effect by tracing back the particles that lie
within the final virial radius to their positions at z=5. Figure 3
shows the fractional mass of the cluster within a given overdensity
contributed by the halos that have collapsed at z=5. In the high
resolution simulation 14\% of the total mass of the final halo has
already collapsed within halos at z=5. This material ends up
constituting 75\% of the mass within $0.01r_{vir}$. Within our medium
resolution simulation of the same halo we find that only 2\% of the
mass of the final halo has collapsed by z=5 and this ends up making
just 20\% of the halo mass within $0.01r_{vir}$. The low resolution
simulation does not contain any virialised halos at $z=5$.

The power spectrum of CDM type models asymptote towards a slope $n=-3$ and
provides power on all scales relevant to cosmological numerical
simulations.  Therefore, most of the material in the CDM universe and
must lie in collapsed virialised clumps at high redshifts.  If the
simulations had even better mass resolution, we would have expected to
observe halos collapsing at much earlier epochs and at higher phase
space densities.  This material would end up at the center of the
final cluster, perhaps altering the density profiles from what we
find at the current resolution.

\section{Summary}

We have performed the highest resolution simulations of cold dark
matter halos to date.  Our force resolution is 0.2\% of the virial
radii and mass resolution is such that we can resolve halos 1/50,000th
of the mass of the final system.  The wealth of substructure within the
final systems is phenomenal and will be the focus of a future paper.
We can identify approximately 1500 ``halos within halo'' at the final
time.  Some of the halos within the cluster's virial radius contain
their own substructure.

Both particle number and softening play a role in shaping the final
density profiles; as we increase the resolution the halo profiles
become steeper and denser in the centers. With the same force softening,
but with 10 times as many particles, an individual halo will be
roughly twice as dense at $0.01r_{vir}$ and increasing the mass
resolution by another order of magnitude leads to densities that are a
further 50\% higher at the same radius. 
Even with 3 million particles per halo we might not have converged
upon a unique density profile for cold dark matter halos. Of order
$10^6$ halo particles must be used in order to compare the dynamics of
the inner few percent of the cluster mass distribution with
observations.


Our convergence tests suggest that our highest resolution simulation
can be used to determine the density profile to a scale just less than
$\sim 1\%$ of the virial radius.  At this point the profile has
reached an asymptotic slope of $r^{-1.4}$. It will be interesting to
study the statistics of gravitational arcs within clusters using
steeper density profiles ({\it c.f.} Bartelmann 1996).  Since galaxy halos will
have steeper profiles at a fixed scale length compared to the cluster
sized halos simulated here, the problem of reconciling cold dark
matter halos with observations of galaxy rotation curves will be
considerably exacerbated by these results.

\acknowledgments

{\bf Acknowledgments} \ \ \ We would like to thank John Dubinski, Gus
Evrard, Carlos Frenk, Neal Katz, Adrian Melott, Julio Navarro and
Simon White for stimulating discussions on halo formation.  We also
thank the referee for comments that helped clarify the paper. Adrian
Jenkins kindly providing the IC's for the cluster comparison run.
This research required a great deal of computational resources and we
are indebted to NASA's High Performance Computing and Communications
program and the Virgo consortium. Computations were performed on the
Pittsburg T3E and the UK CCC Origin 2000. BM is a Royal Society
University Research Fellow, FG is supported by the European Network grant
``Galaxy Formation and Evolution''.

\baselineskip=8pt

\clearpage
\vskip 1.0truein

\noindent{\bf References}



\pp Bartelmann M. 1996, {\it A.A.}, {\bf 313}, 697.

\pp Bertschinger E. 1985, {\it Ap.J.Supp.}, {\bf 58}, 39.

\pp Carlberg R. 1994, {\it Ap.J.}, {\bf 433}, 468.

\pp Couchman H.M.P., 1991, {\it Ap.J.Lett.}, {\bf 368}, 23L.

\pp Craig M. 1997, {Ph.D. thesis}, University of California, Berkeley.

\pp Crone M.M., Evrard A.E. \& Richstone D.O. 1994, {\it Ap.J.}, {\bf 434}, 402.

\pp Davis M., Efstathiou G., Frenk C.S. \& White S.D.M. 1985, 
{Ap.J.}, {\bf 292}, 371.

\pp Dubinski J. \& Carlberg R. 1991, {\it Ap.J.}, {\bf 378}, 496.

\pp Eke V.R., Cole S. \& Frenk C.S. 1996, {\it M.N.R.A.S.}, {\bf 282}, 263.

\pp Evans W.N. \& Collet 1997, {\it Ap.J.Lett.}, in press.

\pp Flores R.A. \& Primack J.R. 1994, {\it Ap.J.Lett.}, {\bf 457}, L5.

\pp Frenk C.S., White S.D.M., Davis M. \& Efstathiou G. 1988, {\it Ap.J.}, 
{\bf 327}. 507.

\pp Frenk C.S., White S.D.M., Jenkins A.R., Pierce F. \& Evrard A. 1998,  
in preparation.

\pp Ghigna S., Moore B., Governato F., Lake G., Quinn T. \& Stadel J.
1998, {\it M.N.R.A.S.}, submitted.

\pp Moore B., 1994, {\it Nature}, {\bf 370}, 620.

\pp Moore B., Katz N. \& Lake G. 1996, {\it Ap.J.}, {\bf 457}, 455.

\pp Navarro J.F., Frenk C.S. \& White S.D.M. 1996, {\it Ap.J.}, {\bf 462}, 563.

\pp Quinn P.J., Salmon J.K. \& Zurek W.H. 1986, {\it Nature}, {\bf 322}, 329.

\pp Subramanian K. \& Ostriker J. {in preparation}

\pp Splinter R.J., Melott A.L. \& Shandarin S.F. 1998, {\it Ap.J.}, in press.

\pp Syer D. \& White S.D.M. 1997, {\it Astro-ph}, 9611065

\pp Tormen G., Bouchet F.R. and White S.D.M. 1996, {\it M.N.R.A.S.}, 
{\bf 286}, 865.

\pp Warren S.W., Quinn P.J., Salmon J.K. \& Zurek H.W. 1992, {\it Ap.J.}, 
{\bf 399}, 405.

\baselineskip=14pt

\clearpage
\centerline{\bf Figure Captions}
\bigskip


\noindent{\bf Colour Plate 1.} \ \ \ The upper panel shows the
particle distribution in the full 1000 Mpc box from the resimulation
of the Coma cluster.  This highlights the scale of the simulation and
the seven refinement zones that focus in on the highest resolution
region.  In the lower panel we show a sphere of radius $r_{vir}$ that
contains the cluster at z=0. All the visible halos and substructure in
this plot lie within the virial radius and the colour scale spans a
factor of $10^5$ in local density.

\noindent{\bf Colour Plate 2.} \ \ \ We show the projected particle
distributions for the Coma cluster at three different resolutions. In
each case, we have extracted a sphere of radius equal to $r_{vir}$ =
3.4 Mpc.  On the left hand side we plot the logarithm of the local
density covering 5 orders of magnitude, from $\rho/\rho_o=10^1 -
10^6$. On the right hand side we plot the local phase space density,
$\rho/\sigma^3$, where $\sigma$ is the local velocity dispersion
calculated by smoothing over the nearest 64 neighbours. 

\noindent{\bf Figure 1.} \ \ \ The density profiles of the Coma
cluster simulated at four different resolutions. The curves begin
at the spline softening lengths that were used and the number of
particles within the final virial radii are indicated.


\noindent{\bf Figure 2.} \ \ \ The broken lines show the Virgo halo
simulated at the same mass resolution but varying only the softening
parameter. This halo has a virial radius of 2 Mpc and contains 20,000
particles within $r_{vir}$.  The values of the softening used are
indicated next to each curve.  The solid curves show the same cluster
resimulated with a mass resolution 20 times higher, but keeping the
force softening fixed at 10kpc.  To demonstrate that relaxation is not
affecting our results one of the solid curves shows the profile at a
redshift z=0.25.



\noindent{\bf Figure 3.} \ \ \ We plot the mass fraction of the final
halo that resides in material identified at z=5 with $\rho/\rho_o>200$,
against the overdensity that that material lies at z=0. The solid
curve denotes our highest resolution simulation of the Coma cluster
and the dashed curve denotes the intermediate resolution
simulation. (The low resolution run did not contain any virialised
halos at z=5.)

\bigskip

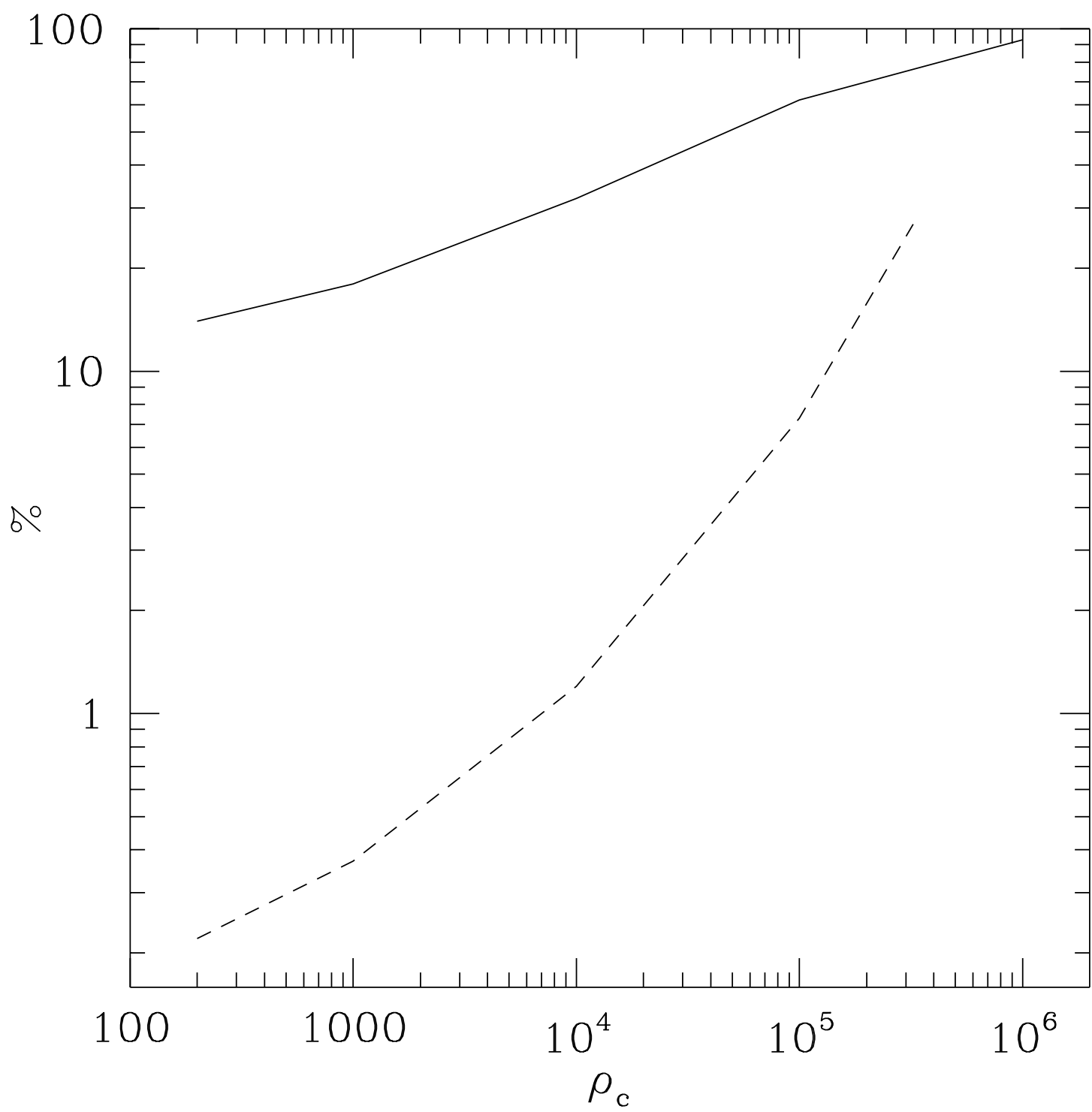
\end{document}